\documentclass[runningheads]{llncs}
\usepackage[T1]{fontenc}
\usepackage{graphicx}
\usepackage[hidelinks]{hyperref}
\usepackage{subcaption}
\usepackage{multirow}

\usepackage{booktabs}
\usepackage{colortbl}
\usepackage{textcomp}
\usepackage{xcolor}
\usepackage{soul}
\usepackage{cite}
\usepackage{balance}
\usepackage{multirow}
\def\BibTeX{{\rm B\kern-.05em{\sc i\kern-.025em b}\kern-.08em
    T\kern-.1667em\lower.7ex\hbox{E}\kern-.125emX}}

\usepackage[acronym,shortcuts]{glossaries-extra}
\glssetcategoryattribute{acronym}{nohyper}{true}
\setabbreviationstyle[acronym]{long-short}
\newacronym{ab}{AB}{AdaBoost}
\newacronym{acn}{ACN}{Adaptive Charging Network}
\newacronym{cps}{CPS}{Cyber-Physical System}
\newacronym{cts}{CTS}{Current Time Series}
\newacronym{dt}{DT}{Decision Tree}
\newacronym{ev}{EV}{Electric Vehicle}
\newacronym{evse}{EVSE}{Electric Vehicle Supply Equipment}
\newacronym{knn}{kNN}{k-Nearest Neighbors}
\newacronym{rf}{RF}{Random Forest}
\newacronym{rn}{ResNet}{Resolution Network}
\newacronym{rsse}{RSSE}{residual sum of squared errors}
\newacronym{soc}{SoC}{State of Charge}
\newacronym{svm}{SVM}{Support Vector Machine}
\newacronym{ts}{TS}{time series}
\newacronym{ada}{ADA}{ADA Boost}
\newacronym{lr}{LR}{Logistic Regression}
\newacronym{li-ion}{Li-ion}{Lithium-ion}
\newacronym{v2g}{V2G}{Vehicle-To-Grid}
\newacronym{ac}{AC}{Alternate Current}
\newacronym{dc}{DC}{Direct Current}
\newacronym{ml}{ML}{Machine Learning}
\newacronym{nn}{NN}{Neural Network}

\newcommand{\rev}[1]{\textcolor{black}{#1}}

\begin{document}
\title{Profiling Electric Vehicles\\via Early Charging Voltage Patterns}
\author{Francesco Marchiori\inst{1}\orcidID{0000-0001-5282-0965} \and
Denis Donadel\inst{2}\orcidID{0000-0002-7050-9369} \and
Alessandro Brighente\inst{1}\orcidID{0000-0001-6138-2995} \and
Mauro~Conti\inst{1}\orcidID{0000-0002-3612-1934}}
\institute{University of Padova, Padova, Italy \and University of Verona, Verona, Italy\\
\email{francesco.marchiori@math.unipd.it, denis.donadel@univr.it, \{alessandro.brighente, mauro.conti\}@unipd.it}}
\maketitle              %
\begin{abstract}
Electric Vehicles (EVs) are rapidly gaining adoption as a sustainable alternative to fuel-powered vehicles, making secure charging infrastructure essential. Despite traditional authentication protocols, 
recent results showed that attackers may steal energy through tailored relay attacks.
One countermeasure is leveraging the EV's fingerprint on the current exchanged during charging. 
However, existing methods focus on the final charging stage, allowing malicious actors to consume substantial energy before being detected and repudiated. 
This underscores the need for earlier and more effective authentication methods to prevent unauthorized charging.
Meanwhile, profiling raises privacy concerns, as uniquely identifying EVs through charging patterns could enable user tracking.

In this paper, we propose a framework for uniquely identifying EVs using physical measurements from the early charging stages. 
We hypothesize that voltage behavior early in the process exhibits similar characteristics to current behavior in later stages. 
By extracting features from early voltage measurements, we demonstrate the feasibility of EV profiling. 
Our approach improves existing methods by enabling faster and more reliable vehicle identification. 
We test our solution on a dataset of 7408 usable charges from 49 EVs, achieving up to 0.86 accuracy. 
Feature importance analysis shows that near-optimal performance is possible with just 10 key features, improving efficiency alongside our lightweight models. 
This research lays the foundation for a novel authentication factor while exposing potential privacy risks from unauthorized access to charging data.

\end{abstract}
\section{Introduction}

The increasing diffusion of \acp{ev}, with total sales forecasted to reach up to 31.1 million by 2030~\cite{deloitte}, is a key factor to help fight global warming. 
At the same time, it opens up to new challenges related to the peculiarity of these devices. 
Many of these challenges are related to the batteries, how to make a full charge last longer, and how to charge them fast and efficiently, while ensuring the grid stability.
The grid leverages communication between vehicles and charging columns by employing the so-called \ac{v2g} paradigm.
Data exchange usually happens using a power line communication through the control pilot pin of the charging plug and is usually enabled by the ISO 15118 protocol~\cite{multin2018iso}.
Through it, \acp{ev} can establish a full-featured internet connection with the smart grid, allowing price negotiation, charging time scheduling, authentication handling, and enabling many other services~\cite{multin2018iso}. While this communication comes with high benefits, it could also be exploited by attackers~\cite{antoun2020detailed, conti2022evexchange, baker2019losing}. 

Being a \ac{cps}, the \ac{ev} charging infrastructure security must not only rely on the protocol's security but also consider threats from the physical world. 
In fact, information leakages can happen through the exchange of physical signals. 
In particular, researchers demonstrated how it is possible to profile an \ac{ev} by looking at the energy delivered by the charging column~\cite{brighente2024evscout2, brighente2021}. This approach collects the charging current in the final part of a full charging process to extract features and characterize each \ac{ev}.
The approach achieves good classification performance and is open to several potential implications, both for attackers who want to trace a vehicle between multiple locations and for defenders as an authentication mechanism. 
However, since the data are collected during the last part of the charging process (the so-called \textit{tail}), the applicability of this approach is limited to charging processes not reaching 100\% of \ac{soc}, thus potentially missing several profiling chances and making authentication more complicated.     

\paragraph{Contributions.}
In this paper, we analyze a different solution that allows \ac{ev} profiling by collecting measurements during the early stages of the charging process.
In particular, we employ voltage measurements, which are significant in the first part of the charging process, instead of current values, which provide more information in the last part of the charging. 
Moreover, while similar works consider a dataset coming from a series of charging columns employing an adaptive charging solution (ACN Dataset~\cite{Lee2019acndata, Lee2019acnsim}), in this paper, we consider the EVBattery dataset~\cite{zhang2023realistic, he2022evbattery}, which provides a more standard and spread charging management system. 
Overall, our findings demonstrate that \ac{ev} authentication can be reliably performed using simple and lightweight models.
Our approach achieves an accuracy of up to 0.86, averaged across all vehicle types.

The contribution of this paper can be summarized as follows:
\begin{itemize}
    \item We analyze the feasibility of profiling a vehicle through measurements of the voltage exchange during the first charging steps. 
    \item We analyze the performance of our framework on a large real-world dataset~\cite{zhang2023realistic}, comprising charges from more than $49$ \acp{ev} and including three different brands.
    Our approach is tested on anonymized data, demonstrating its robustness in privacy-preserving scenarios.
    \item We identify the most relevant features for \ac{ev} authentication, showing that a small subset of 10 key features is sufficient for high accuracy.
    \item We make our implementation and code open-source at: \url{https://github.com/spritz-group/EV-Volt-Auth}.
\end{itemize}

\paragraph{Organization.}
This paper is organized as follows.
In Section~\ref{sec:realted} we report related works in the field of \ac{ev} security and authentication.
We then detail the considered system and threat model in Section~\ref{sec:system}, and in Section~\ref{sec:methodology} we propose our methodology.
We report the results of our evaluation in Section~\ref{sec:evaluation}, and Section~\ref{sec:conclusions} concludes our work.

\section{Related Works}
\label{sec:realted}

Several research efforts have explored the security and privacy aspects of \acp{ev}~\cite{antoun2020detailed, gottumukkala2019cyber, muhammad2023emerging}.
Looking at the bigger picture, the need for \ac{ev} to periodically connect to the grid for charging can threaten the entire power grid. Its stability could be mined by botnets of vehicles~\cite{khan2019impact} and by coordinated attacks~\cite{ghafouri2022coordinated}. 
Moreover, the backend connection between control centers and the charging columns opens up several challenges~\cite{nasr2023chargeprint}. For instance, the most widely used protocol for backend communications has been proved vulnerable to cyberattacks~\cite{alcaraz2023ocpp}.

Being a \ac{cps}, the \ac{ev} itself exhibits different security risks in addition to traditional petrol-powered cars~\cite{ye2020cyber, brighente2023electric}.
Attackers may exploit the physical connection to the charging columns to steal energy from a nearby victim~\cite{conti2022evexchange} or perform denial of charge attacks, even remotely~\cite{baker2019losing}.
Different authentication strategies have been proposed to secure the charging process, investigating also the dynamic charging that allows cars to charge while moving~\cite{li2016portunes+, babu2021robust, mookherji2024secure}.
Brighente at al.~\cite{brighente2021} first introduced an approach for authentication by employing information from the \ac{ev} specific charging pattern, which was then expanded in different following works~\cite{gangwal2023feasibility, brighente2024evscout2}.
However, their approach requires a full charge of the \ac{ev} battery before producing a result, thus reducing the possible applicability of the framework.
The same issue holds for authentication methods aimed at lithium-ion batteries, as they can require full battery charges and discharges or sophisticated equipment, making their usability in the context of \acp{ev} limited~\cite{marchiori2023your}.

Privacy aspects of \ac{ev} charging have also been investigated~\cite{unterweger2022analysis}.
However, it is not always easy to prevent \ac{cps} from leaking information that an attacker may exploit to mine the user's privacy~\cite{zhang2016privacy}.
Recent work has demonstrated the feasibility of extracting sensitive information—such as user identity, driving style, and trip endpoints purely from battery consumption patterns using \ac{ml} techniques~\cite{marchiori2025leaky}.
While this highlights the privacy risks inherent in battery-related telemetry, such approaches focus on post-drive consumption data, typically requiring access to a full trip.
In contrast, our work shifts the focus to the early stages of the charging process, showing that brief voltage readings alone can be leveraged for profiling, even without full charging sessions or vehicle usage data.
This significantly broadens the threat landscape, as it reduces the time and access requirements for potential adversaries.

\section{System and Threat Model}
\label{sec:system}

In this section, we describe the system under study (Section~\ref{subsec:system}), outline the adversarial model (Section~\ref{subsec:threat}), and explore potential use cases of our profiling method (Section~\ref{subsec:applications}).
We distinguish between malicious and legitimate applications, highlighting the dual-use nature of \ac{ev} fingerprinting via voltage traces.

\subsection{System Model}
\label{subsec:system}

Charging an \ac{ev} requires careful consideration of both the vehicle and the grid.
This latter, in particular, must be managed to handle high power requirements from a fleet of \acp{ev}, and this is facilitated by the \ac{v2g} paradigm that creates a bi-directional communication between vehicles and the smart grid.
On the other side, \ac{ev}'s battery can be charged at different speeds based on the capabilities of the charging column and the vehicle itself~\cite{khalid2021comprehensive}. 

\begin{figure}[t]
    \centering
    \includegraphics[width=.75\columnwidth]{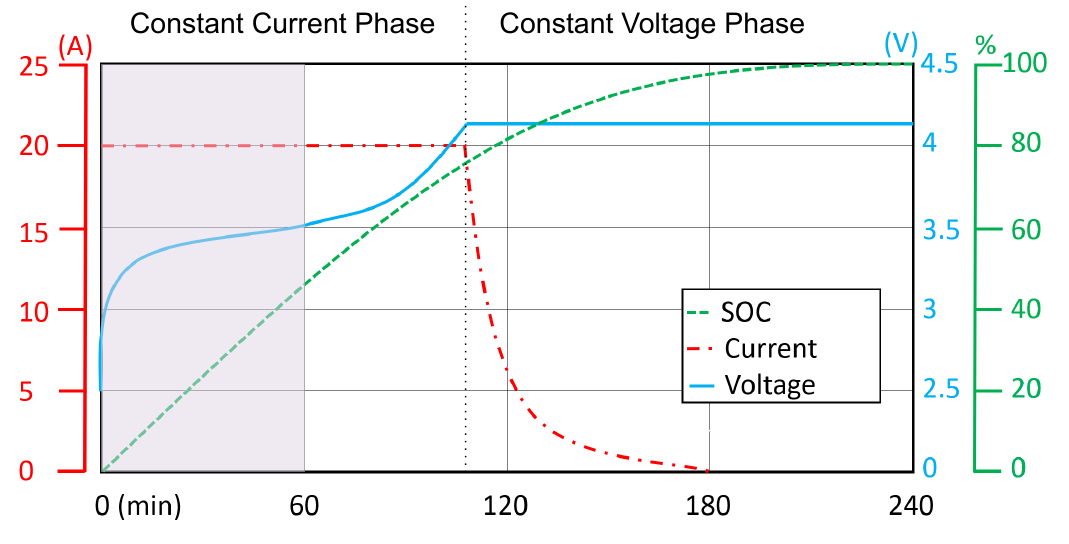}
    \caption{Charging profile of a Li-ion battery~\cite{thundersky}. In this paper, we will show that voltage data from the first part of the constant current phase (highlighted in violet in the graph) are enough for correctly profiling a vehicle.}
    \label{fig:soc}
\end{figure}

\acp{ev} are usually equipped with batteries containing \ac{li-ion} cells. 
They exhibit a known and peculiar charging pattern where voltage and current levels are correlated to the \ac{soc}. 
In particular, two main behaviors can be identified, as shown in Fig.~\ref{fig:soc}. 
The charging starts with a \textit{constant current phase} where the voltage slowly increases. 
At a certain point, usually with \ac{soc} values between 60\% and 80\% of the full charge, a \textit{constant voltage phase} starts, where the current slowly decreases, creating a tail. 
This behavior is well known and is typical of these kinds of batteries~\cite{shen2012charging}.

\subsection{Threat Model}
\label{subsec:threat}

In this paper, we consider an entity that can collect measurements during the charging process. 
In particular, with respect to previous works~\cite{brighente2024evscout2} that collected current levels, we collect voltage levels during the charging process.
As we discuss in this paper, this reduces the required collection time to profile a vehicle and does not require a full charge to extract the relevant features. 
The only other information required by the system is the \ac{soc} of the battery, which is used to understand in which part of the charging profile the \ac{ev} is. 
All the other data transmitted through the cable, such as the high-level communication transmitted in the control pilot pin, is not collected for this study.  

Such profiling could be a double-edged sword and can be applied by different entities for different purposes. 
An attacker may exploit this methodology to track a vehicle through different locations. 
A malicious owner of various parking lots with charging columns may track vehicles for advertisement purposes. 
Another option for an attacker is to install a measurement device as a plug, similar to devices employed in ATM skimming~\cite{atmSkimming}, to collect data useful to track a user mining their privacy. 
On the other side, such data could be employed for good as a second factor or continuous authentication system. 

\subsection{Applications}
\label{subsec:applications}

We show that it is possible to extract discriminative features from early charging voltage patterns, which are unique enough to allow reliable profiling of a singular vehicle.
The ability to fingerprint an \ac{ev} from voltage measurements can be leveraged in multiple contexts.
We highlight both adversarial and defensive uses.

\paragraph{Adversarial Applications.}
\begin{itemize}
    \item \textbf{Vehicle tracking across locations:} A malicious entity operating multiple public or semi-public charging stations can silently collect voltage traces and match them across sites to track vehicles over time.
    \item \textbf{Profiling without consent:} An attacker could embed a skimming device within a charging cable or socket to passively collect voltage traces and associate them with a specific vehicle or user, compromising location privacy. \item \textbf{Behavioral surveillance:} By linking profiling data with usage patterns, an attacker might infer sensitive behavior (e.g., commuting habits, work location, or home address).
\end{itemize}

\paragraph{Defensive Applications.}
\begin{itemize}
    \item \textbf{Second-factor authentication:} \ac{ev} profiling can supplement existing user authentication mechanisms, confirming vehicle identity as part of a multifactor security scheme at high-security charging stations.
    \item \textbf{Tamper detection:} Deviations in the expected voltage pattern could be used to detect unauthorized battery replacement or tampering with internal battery components.
    \item \textbf{Anti-theft tracking:} In case of theft, a known voltage fingerprint can serve as a unique signature to detect the vehicle if it is connected to any charging infrastructure.
    \item \textbf{Usage-based insurance or leasing:} Insurance or leasing companies may use this approach to verify the \ac{ev}'s identity under a specific contract, enforcing user-vehicle binding in shared or rental contexts.
\end{itemize}

These use cases demonstrate that early-stage voltage profiling has a wide range of implications, from privacy threats to enabling lightweight security mechanisms.
Our work does not advocate for any particular application but aims to provide a technical foundation and empirical validation for \ac{ev} identification based on early charging behavior.
Depending on who controls the charging infrastructure, this capability may pose a risk to user privacy or be a valuable tool for improving \ac{ev} security and authentication.
\section{Methodology}
\label{sec:methodology}

\begin{figure}[!h]
    \centering
    \includegraphics[width=.8\linewidth]{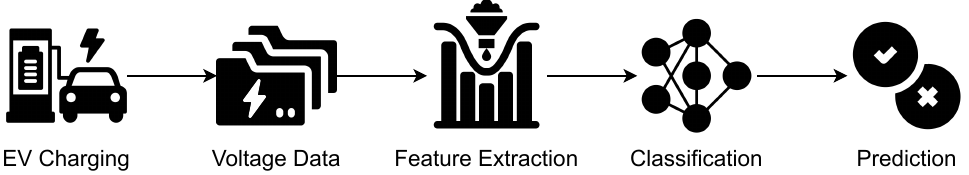}
    \caption{Pipeline of the proposed profiling framework.}
    \label{fig:schema}
\end{figure}

In this section, we provide a more detailed explanation of the \rev{profiling} %
techniques we employ.
We first present the dataset and its characteristics (Section~\ref{subsec:dataset}).
We also discuss our feature extraction process, which is one of the key components of our approach (Section~\ref{subsec:features}). 
Next, we present an overview of the \ac{ml} models used as classifiers and the process of finding their optimal hyperparameters (Section~\ref{subsec:models}).
The overall methodology is summarized in Fig.~\ref{fig:schema}. The first step is collecting charging measurements from \ac{ev} charging process. From the samples, feature extraction is used to obtain information that can be fed to \ac{ml} models for classification.

\subsection{Dataset}
\label{subsec:dataset}

For this work, we created a dataset starting from three \ac{ev} charging collections, which together consist of over 690,000 charging snippets recorded from 347 distinct \acp{ev}~\cite{zhang2023realistic}.
Since the original dataset was designed for anomaly detection, we first remove all potential outliers that exhibit anomalies in the charging process.
Next, we extract all charging snippets, which originally are segmented into 128-sample sequences, with each sample recorded once per second.
Our primary feature of interest is voltage, though we also incorporate \ac{soc} for further processing.
To ensure we analyze complete charging events, we concatenate snippets belonging to the same charging session, grouping them based on the car label and segment ID.
This process results in complete charge sequences, which we sort by \ac{soc} to maintain its natural increasing order.
After preprocessing, our dataset consists of voltage and \ac{soc} time series, and car labels. From this, we apply two additional filtering steps.
\begin{itemize}
    \item \textit{\ac{soc} Thresholding} -- We keep only data where \ac{soc} is  $\leq60\%$, as our \rev{profiling} approach relies on the non-constant voltage phase in the early charging stages (see Fig.~\ref{fig:soc}).
    \item \textit{Minimum Sample Requirement} -- We include only vehicles with at least 100 charging samples, ensuring sufficient data for profiling.
    Cars with few charging instances are excluded due to insufficient label representation in the dataset.
\end{itemize}

\begin{figure}[!h]
    \centering
    \includegraphics[width=.7\linewidth]{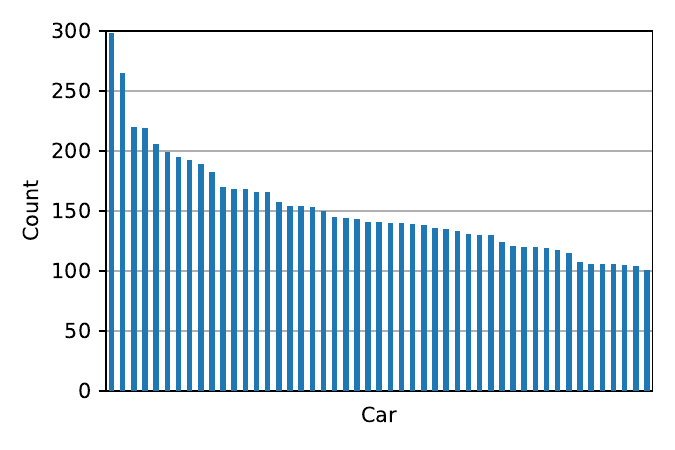}
    \caption{Distribution of samples for each car in our dataset after pre-processing.}
    \label{fig:dataset}
\end{figure}

The final dataset thus contains 36,165 charging snippets constituting 7,408 charging sessions from 49 different \acp{ev}.
It is worth noting that the voltage data has been perturbed and interpolated as part of the anonymization process applied by the original dataset authors~\cite{zhang2023realistic}.
Despite these modifications, underlying temporal patterns and correlations remain intact, potentially enabling the inference of an \ac{ev}'s identity, an aspect we explore further in Section~\ref{subsec:anon}.
After this one-time preprocessing step, we further adapt the dataset dynamically to be used in a classification setup.
The final distribution of the dataset is shown in Fig.~\ref{fig:dataset}.
From now on, we will discuss an authentication scenario, but a malicious user could apply the same process to profile and track a vehicle between charging stations.
The only subtle difference is that an attacker may not have access to detailed \ac{soc} data since the information could be transmitted encrypted or on channels not under the attacker's control. We will discuss this issue in Section~\ref{subsec:soc}.

In particular, we set up our systems to first select a vehicle as the authenticated subject for each car label and consider all other vehicles as non-authenticated.
This process is repeated for every vehicle in the dataset, ensuring that each car is evaluated as the authenticated one at least once.
By structuring the dataset this way, we create multiple binary classification tasks tailored to a specific vehicle's charging behavior.

\subsection{Feature Extraction}
\label{subsec:features}

For each time window, we extract features from the voltage and \ac{soc} time series using the \texttt{tsfresh} Python package~\cite{tsfresh}.
\texttt{tsfresh} automatically computes a wide range of statistical and mathematical characteristics from time series data and selects the most relevant ones for regression or classification tasks.
The extracted features include but are not limited to, statistical measures (e.g., mean, variance, skewness, kurtosis), frequency-domain features (e.g., Fourier coefficients, spectral entropy), and time-series properties (e.g., autocorrelation, trend strength, peaks, crossings).
We apply this process to the voltage and \ac{soc} time series from our processed dataset, ensuring that each driver performs feature extraction separately.
This allows us to generate feature representations tailored to each driver’s charging behavior, which we then use in a binary classification setting (authenticated vs. non-authenticated).
After extracting features, we perform two processing steps.
\begin{enumerate}
    \item \textit{Feature Imputation} – Some extracted features may contain missing values due to insufficient data in certain time windows.
    We handle this by imputing missing values, typically by replacing them with appropriate statistical estimates (e.g., mean, median, or interpolation methods), ensuring a complete dataset for classification.
    \item \textit{Feature Selection} – Since \texttt{tsfresh} generates many features, we apply feature selection to retain only the most informative and discriminative ones.
    The reduction of feature number is done independently for each \ac{ev} to maintain the most suited features in each case.
    This step reduces dimensionality and improves model efficiency, removing noisy or redundant features.
    Although feature selection is performed independently for each vehicle, we observe a significant overlap in the selected features across different models.
    This suggests that certain statistical patterns are consistently informative, regardless of the individual EV's characteristics.
    We further investigate this in Section~\ref{subsec:features-analysis}, where we analyze feature importance and demonstrate that a small set of just 10 common features is sufficient to achieve performance close to that of the full feature set.
\end{enumerate}

\subsection{Models}
\label{subsec:models}

Inspired by previous work on \ac{ev} authentication~\cite{brighente2024evscout2} and battery authentication~\cite{marchiori2023your}, we select five lightweight \ac{ml} models: AdaBoost, \ac{dt}, \ac{knn}, \ac{nn}, and \ac{rf}.
These models were chosen not only based on their effectiveness in prior studies but also due to their suitability for real-time deployment.
Unlike deep learning approaches, which typically require high computational power and external data processing, these models are lightweight enough to run directly on an \ac{ev} charging station without requiring communication with external servers, enhancing security by reducing potential attack vectors associated with remote authentication.
We apply an 80/20 split between the training and test sets for model training.
Hyperparameter tuning is performed using a grid search approach with 5-fold cross-validation, ensuring that our models generalize well to unseen data.
The set of hyperparameters considered for each model is detailed in the Appendix~\ref{app:hyperparams}. %

\section{Evaluation}
\label{sec:evaluation}

We now proceed with the evaluation of our models under different settings. In particular, we analyzed the behavior our our models when changing the measurement length to extract the features from (Section~\ref{subsec:length}) and when varying the composition of the dataset (Section~\ref{subsec:ratio}). Moreover, we analyzed the importance of the \ac{soc} feature in Section~\ref{subsec:soc}, the features employed and the effects on reducing the feature number in Section~\ref{subsec:features-analysis}.

To evaluate our models, we employ two classical metrics, the accuracy and the F1-score, which are defined as follows: 

\begin{equation}
    Accuracy = \frac{TP + TN}{TP + FP + TN + FN}, \quad\quad
    F1 = \frac{2TP}{2TP + FP + FN},
\end{equation}
where: 
\begin{itemize}
    \item \emph{True Positive (TP)}: a sample from the authenticated vehicle is correctly identified.
    \item \emph{False Positive (FP)}: a sample belonging to the non-authenticated class mistakenly identified as the authenticated vehicle.
    \item \emph{False Negative (FN)}: a sample belonging to the authenticated driver is mistakenly identified as a non-authenticated driver.
    \item \emph{True Negative (TN)}: a non-authenticated vehicle is correctly classified as non-authenticated.
\end{itemize}

\subsection{Data Length}
\label{subsec:length}

\begin{figure*}[t]
    \centering
    \begin{subfigure}{0.7\textwidth}
        \centering
        \includegraphics[width=.925\textwidth]{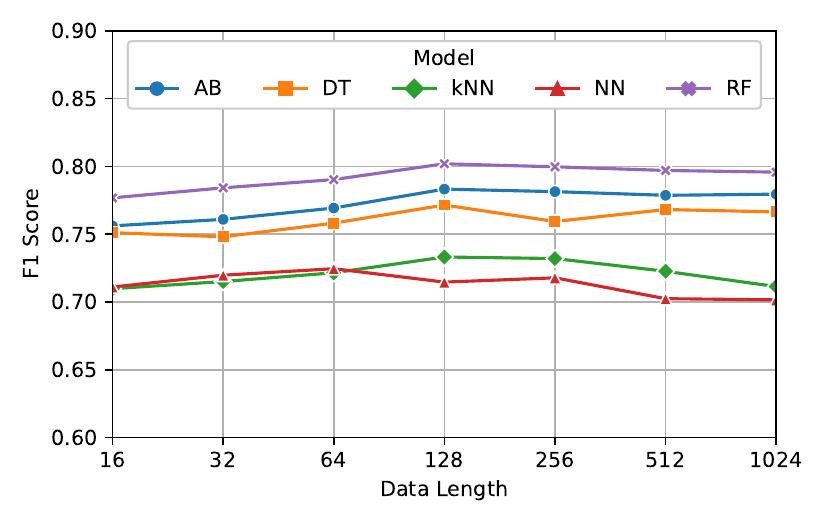}
        \caption{F1 Score of the models.}
        \label{subfig:res_length_f1}
    \end{subfigure}
    \begin{subfigure}{0.7\textwidth}
        \centering
        \includegraphics[width=.925\textwidth]{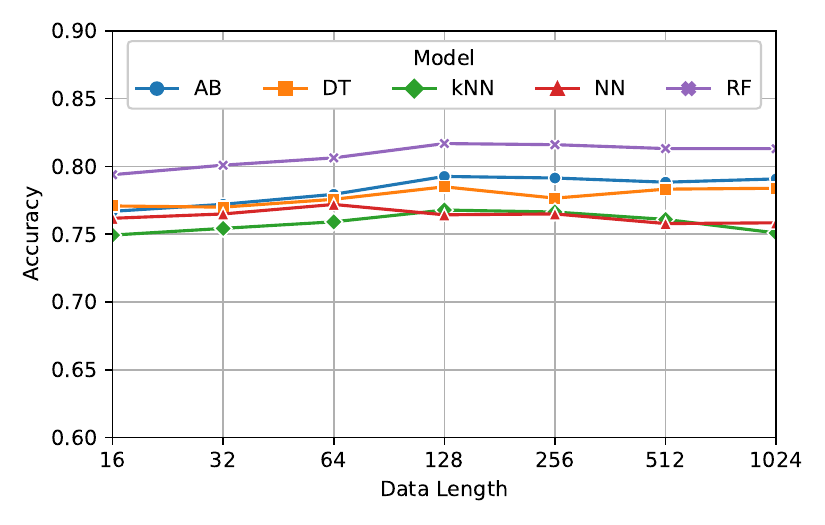}
        \caption{Accuracy of the models.}
        \label{subfig:res_length_accuracy}
    \end{subfigure}
    \caption{Average performance of different data lengths (in seconds).}
    \label{fig:length}
\end{figure*}

One of the most important differences with respect to previous works~\cite{brighente2024evscout2} is that our models do not require a full charge to efficiently profile a vehicle. In fact, since we extract features from the first part of the charging, only a few seconds of charging are sufficient to obtain some results. To understand how much the size of the data capture has an impact on the performance, we tested several options ranging from 16 to 1024 seconds.

Results are shown in Fig.~\ref{fig:length}, where Fig.~\ref{subfig:res_length_f1} shows the F1-scores, while Fig.~\ref{subfig:res_length_accuracy} the accuracy, both averaged by the ratios (see Section~\ref{subsec:ratio}). 
As we can see, the overall variance between sizes is not very pronounced, and this suggests it is not necessary to wait long to make a decision. 
In particular, top scores are reached with a 128-second length by the \ac{rf} classifier, which overall performed quite better than other models.
Another interesting aspect is related to the results of different models within the same data length. 
The \ac{rf} models retain an almost constant distance from the second most performing model over all the data lengths.
Although \ac{knn} and \ac{nn} represent two very different architectures, they share a curve that sees rising scores up to mid-length (i.e., 64 to 128) and then falling again as length increases.

\subsection{Ratio}
\label{subsec:ratio}
We conduct several experiments to verify the effect of an unbalanced dataset during training. To analyze it, we tested our models based on different ratios between the data concerning the positive label versus all the others. In particular, a ratio of $5:1$ indicates that the samples belonging to the selected \ac{ev} are five times more than samples from all the other vehicles. Conversely, a ratio of $1:5$ indicates that the dataset contains five times more samples of other vehicles with respect to the selected one. A ratio of $1:1$ indicates a balanced dataset.  

\begin{figure*}[t]
    \centering
    \begin{subfigure}{0.7\textwidth}
        \centering
        \includegraphics[width=.925\textwidth]{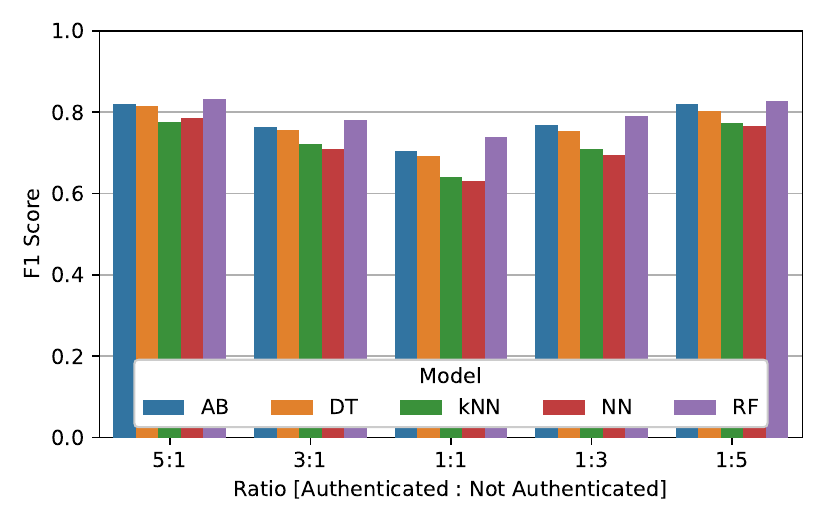}
        \caption{F1 Score of the models.}
        \label{subfig:ratio_f1}
    \end{subfigure}
    \begin{subfigure}{0.7\textwidth}
        \centering
        \includegraphics[width=.925\textwidth]{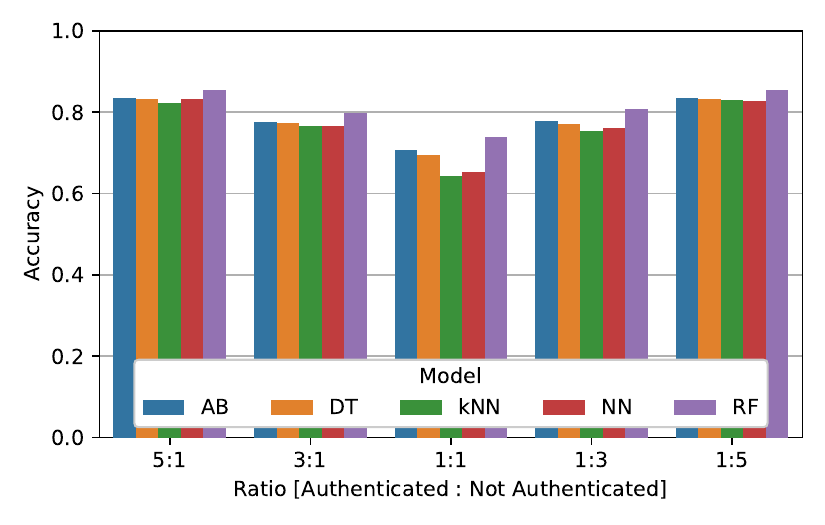}
        \caption{Accuracy of the models.}
        \label{subfig:ratio_accuracy}
    \end{subfigure}
    \caption{Average performance of different ratios (representing the number of authenticated samples vs the number of other samples in the dataset).}
    \label{fig:ratio}
\end{figure*}

The results shown in Fig.~\ref{fig:ratio} clearly indicate unbalanced datasets' capabilities to perform successfully. 
This is essential in such a scenario where it is usually more difficult to get samples from the authenticated vehicle with respect to data from all the other cars that use the charging column. In particular, we can see how with one-fifth of data related to the target (i.e., ratio $1:5$), the F1-score surpasses 0.80 for three models out of five. 
Even in this experiment, \ac{rf} exhibits the best performances. However, the gap is more pronounced with a balanced dataset, while \ac{ab} almost reaches the F1-Score levels of \ac{rf} for heavy unbalanced datasets.

\subsection{\ac{soc} Feature}
\label{subsec:soc}

\begin{figure}[!htbp]
    \centering
    \includegraphics[width=.65\columnwidth]{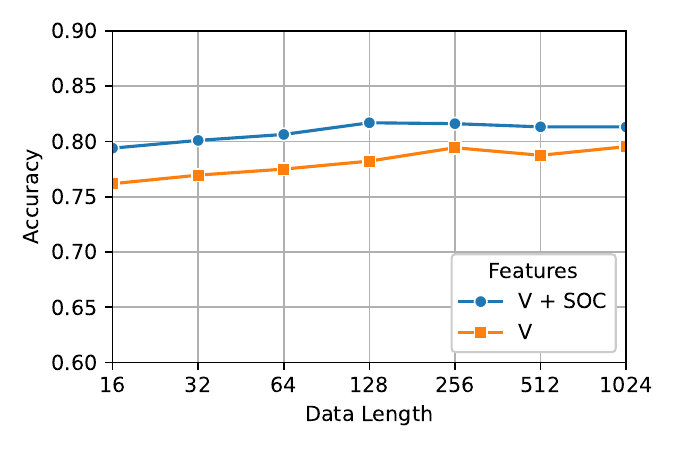}
    \caption{Average performance of our \ac{rf} model trained on different subsets of features (\textbf{V}oltage alone or with \textbf{\ac{soc}})  at different data lengths.}
    \label{fig:comparison}
\end{figure}

In our experiments we employ the \ac{soc} as a time series for feature extraction through \texttt{tsfresh}. This is reasonable since we discussed an authentication scenario where this information should be available to the charging column with a high level of detail.
However, in an adversarial scenario, an attacker may only be able to get rough information about the \ac{soc} for instance by looking at the charging column display, or by estimating it from the voltage level and the duration of the charging. 
In this scenario, the attacker could successfully identify if the charging is happening during the constant current phase (see Fig.~\ref{fig:soc}), but cannot use the \ac{soc} to extract features.
To determine the impact of the absence of the \ac{soc} feature, we reproduced our experiments without considering it during the feature extraction, but employing it only to filter out samples with more than 60\% of \ac{soc}.
The results of this analysis are shown in Fig.~\ref{fig:comparison}.

As we can see, results are comparable, although the presence of \ac{soc}-related features slightly improve the performances of our classifier.
Interestingly, we notice a more significant gap in accuracy performance as the considered data length for classification decreases.
This suggests that \ac{soc}-related features help the model more accurately assess the specific charging phase.
However, the model can infer this independently with more considerable data lengths, as suggested by the smaller gap between the accuracy results.

\subsection{Feature Analysis}\label{subsec:features-analysis}

A complete analysis of the features extracted and employed by \texttt{tsfresh} is useful to understand which trait of the charging is more characteristic and thus useful in the profiling. 
We performed our feature analysis on a subsample of the whole datasets and employed the best-performing model, i.e., the \ac{rf} model. 

\begin{figure}[!htbp]
    \centering
    \includegraphics[width=.7\columnwidth]{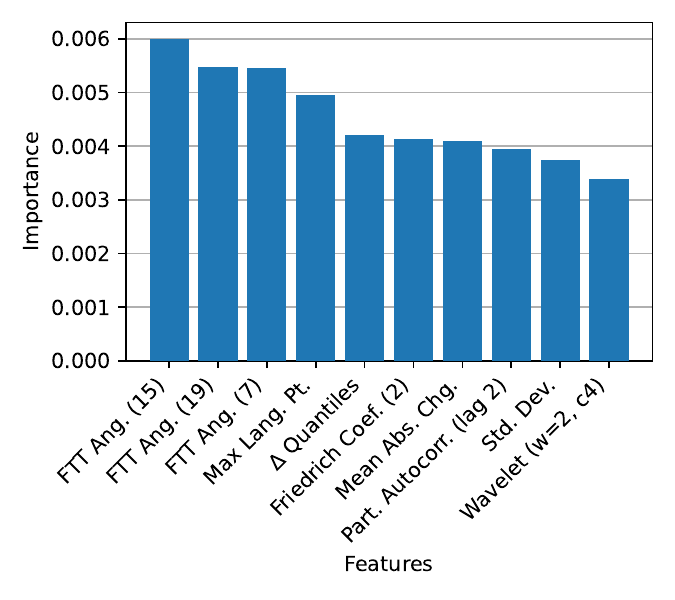}
    \caption{Top 10 features by importance extracted by SHAP.}
    \label{fig:importance}
\end{figure}

We start by analyzing the feature's importance by employing SHapley Additive exPlanations (SHAP), a model-agnostic XAI technique~\cite{NIPS2017_7062}.
Fig.~\ref{fig:importance} classify the main features with their importance level, which are explained more into detail in Appendix~\ref{app:features}.
The prominence of frequency-domain features among the top 10 highlights the significance of periodic voltage signal behavior in \ac{ev} authentication.
The inclusion of both low- and high-frequency components suggests that variations in power electronics and charging circuit dynamics are distinctive to individual \acp{ev}.
Additionally, the presence of features such as mean absolute change and standard deviation indicates that, while absolute voltage levels may be comparable across \acp{ev}, their dynamic fluctuations provide a unique signature for identification.

\begin{figure}[!htbp]
    \centering
    \includegraphics[width=.65\columnwidth]{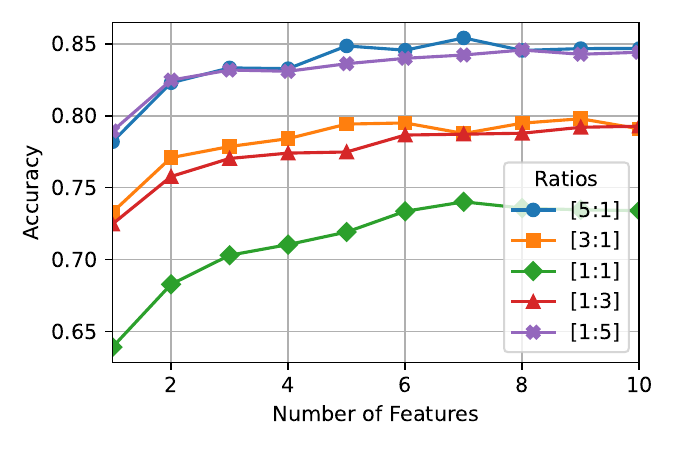}
    \caption{Accuracy of the classifiers trained on different dataset ratios (i.e., number of authenticated samples vs number of other samples) while increasing the number of features. }
    \label{fig:features-number}
\end{figure}

After understanding the most significant features, another important aspect is the impact of adding features on the scores.
A low number of features reduces the complexity of the model, making it more easy to handle and less exposed to noise.
Therefore, we experiment with different numbers of features in a \ac{rf} model, choosing every time the $x$ most significant features.
Results are shown in Fig.~\ref{fig:features-number}.
They show that increasing the number of features generally improves accuracy, but the rate of improvement diminishes as more features are added, becoming almost negligible from 6 features on.
Moreover, the effect varies depending on the data balance ratio.
Higher imbalance ratios tend to achieve better performance overall, suggesting that the model benefits from more training samples in the majority class.
Even with balanced datasets, accuracy improves until plateauing, showing that a feature subset can achieve near-optimal results while keeping the model efficient.

\subsection{Data Anonymization}
\label{subsec:anon}

The authors of the original dataset~\cite{zhang2023realistic} applied a series of anonymization steps to protect the identity and characteristics of the \acp{ev}.
Specifically, they perturbed and interpolated each charging segment's average voltage, current, and temperature values. 
Additionally, timestamps and mileage values were randomly shifted and scaled.
While these anonymization techniques help mask exact values and protect sensitive identifiers, our results show that meaningful patterns remain.
In particular, the temporal correlations and statistical structures in the voltage data can effectively be leveraged to infer the \ac{ev}'s identity.
This demonstrates the robustness of our methodology not only in idealized settings but also when applied to anonymized real-world data.
This highlights the effectiveness of our methodology in both system and threat model scenarios.
\begin{itemize}
    \item \textit{Benign use case:} When employed for authentication at the charging station level, our method enables privacy-preserving user recognition resilient against side-channel attacks on battery usage~\cite{marchiori2025leaky}.
    \item \textit{Adversarial use case:} When used for malicious profiling, our findings indicate that the current anonymization strategy may be insufficient, pointing to stronger or alternative techniques to prevent identity inference.
\end{itemize}

\section{Conclusions}
\label{sec:conclusions}

In this paper, we analyzed the feasibility of profiling an \ac{ev} during charging by exploiting voltage measurements and the \ac{soc}. The data employed are collected during the first part of the charging process, thus allowing vehicle identification during the first stages of the charging, without the need to wait for full charging (i.e., 100\% \ac{soc}). Our experiments on 49 different vehicles show accuracy up to 0.86, demonstrating the practicality of the approach.

Privacy is an essential requirement in the \ac{cps} environment. Looking at potential leakages in the \ac{ev} scenario is important to defend users against possible misuse of such information. Moreover, in this context, this information could also be employed to enhance the security of the system, providing novel features for authentication. In fact, we foresee future work that can build a fully featured second-factor and continuous authentication system starting from this paper's findings.
Using advanced preprocessing can also enhance scores and enable more accurate, robust profiling. Moreover, a larger dataset could allow for investigating the resiliency of the solution to physical properties such as battery aging, seasonality, or time of day. 

\subsection*{Open Science and Ethical Considerations}

This work builds upon publicly available, pre-processed \ac{ev} charging data released for research purposes.
While raw data remain protected under privacy regulations, the processed dataset includes vehicle identifiers that allow charging sessions to be linked to specific (anonymized) \acp{ev}.
Our findings demonstrate that even with such limited information, accurate profiling remains feasible---highlighting a broader security and privacy consideration for the community.
We believe it is important to share these results to help guide the development of more privacy-preserving data-sharing practices and protocols in future \ac{ev} infrastructures.
All our code is available to promote transparency and reproducibility: \url{https://github.com/spritz-group/EV-Volt-Auth}.

\subsection*{Acknowledgment}
This work was founded by the European Union under the National
Recovery and Resilience Plan (NRRP), Mission 4 Component 2 Investment 1.3 - Call for tender No. 341 of March
15, 2022 of Italian Ministry of University and Research –
NextGenerationEU; Code PE00000014, Concession Decree
No. 1556 of October 11, 2022 CUP D43C22003050001,
Project ``SEcurity and RIghts in the CyberSpace (SERICS) -
Spoke 7 Infrastructure Security - Visible Light Communication
for Secure Vehicle-to-Everything Communication - VisiCar'' –
Beneficiary’s CUP: C99J24000250008.
\bibliographystyle{splncs04}
\bibliography{bibliography}

\begin{thebibliography}{10}
\providecommand{\url}[1]{\texttt{#1}}
\providecommand{\urlprefix}{URL }
\providecommand{\doi}[1]{https://doi.org/#1}

\bibitem{alcaraz2023ocpp}
Alcaraz, C., Cumplido, J., Trivino, A.: Ocpp in the spotlight: threats and countermeasures for electric vehicle charging infrastructures 4.0. International Journal of Information Security  \textbf{22}(5),  1395--1421 (2023)

\bibitem{antoun2020detailed}
Antoun, J., Kabir, M.E., Moussa, B., Atallah, R., Assi, C.: A detailed security assessment of the ev charging ecosystem. IEEE Network  \textbf{34}(3),  200--207 (2020)

\bibitem{babu2021robust}
Babu, P.R., Amin, R., Reddy, A.G., Das, A.K., Susilo, W., Park, Y.: Robust authentication protocol for dynamic charging system of electric vehicles. IEEE Transactions on Vehicular Technology  \textbf{70}(11),  11338--11351 (2021)

\bibitem{baker2019losing}
Baker, R., Martinovic, I.: Losing the car keys: Wireless {PHY-Layer} insecurity in {EV} charging. In: 28th USENIX Security Symposium (USENIX Security 19). pp. 407--424. USENIX Association, Santa Clara, CA (Aug 2019), \url{https://www.usenix.org/conference/usenixsecurity19/presentation/baker}

\bibitem{brighente2023electric}
Brighente, A., Conti, M., Donadel, D., Poovendran, R., Turrin, F., Zhou, J.: Electric vehicles security and privacy: Challenges, solutions, and future needs. arXiv preprint arXiv:2301.04587  (2023)

\bibitem{brighente2024evscout2}
Brighente, A., Conti, M., Donadel, D., Turrin, F.: Evscout2. 0: Electric vehicle profiling through charging profile. ACM Transactions on Cyber-Physical Systems  \textbf{8}(2),  1--24 (2024)

\bibitem{brighente2021}
Brighente, A., Conti, M., Sadaf, I.: Tell me how you re-charge, i will tell you where you drove to: Electric vehicles profiling based on charging-current demand. In: Bertino, E., Shulman, H., Waidner, M. (eds.) Computer Security -- ESORICS 2021. pp. 651--667. Springer International Publishing, Cham (2021)

\bibitem{tsfresh}
Christ, M., Braun, N., Neuffer, J., Kempa-Liehr, A.W.: {Time Series FeatuRe Extraction on basis of Scalable Hypothesis tests (tsfresh – A Python package)}. Neurocomputing  \textbf{307},  72--77 (2018). \doi{https://doi.org/10.1016/j.neucom.2018.03.067}

\bibitem{conti2022evexchange}
Conti, M., Donadel, D., Poovendran, R., Turrin, F.: Evexchange: A relay attack on electric vehicle charging system. In: European Symposium on Research in Computer Security. pp. 488--508. Springer (2022)

\bibitem{deloitte}
Deloitte: Electric vehicles: setting a course for 2030. In: https://www2.deloitte.com/us/en/insights/focus/future-of-mobility/electric-vehicle-trends-2030.html (Jul 2020)

\bibitem{gangwal2023feasibility}
Gangwal, A., Jain, A., Conti, M., et~al.: On the feasibility of profiling electric vehicles through charging data. In: Inaugural International Symposium on Vehicle Security \& Privacy (2023)

\bibitem{ghafouri2022coordinated}
Ghafouri, M., Kabir, E., Moussa, B., Assi, C.: Coordinated charging and discharging of electric vehicles: A new class of switching attacks. ACM Transactions on Cyber-Physical Systems (TCPS)  \textbf{6}(3),  1--26 (2022)

\bibitem{gottumukkala2019cyber}
Gottumukkala, R., Merchant, R., Tauzin, A., Leon, K., Roche, A., Darby, P.: Cyber-physical system security of vehicle charging stations. In: 2019 IEEE Green Technologies Conference(GreenTech). pp.~1--5 (2019)

\bibitem{he2022evbattery}
He, H., Zhang, J., Wang, Y., Jiang, B., Huang, S., Wang, C., Zhang, Y., Xiong, G., Han, X., Guo, D., et~al.: Evbattery: A large-scale electric vehicle dataset for battery health and capacity estimation. arXiv preprint arXiv:2201.12358  (2022)

\bibitem{khalid2021comprehensive}
Khalid, M.R., Khan, I.A., Hameed, S., Asghar, M.S.J., Ro, J.: A comprehensive review on structural topologies, power levels, energy storage systems, and standards for electric vehicle charging stations and their impacts on grid. IEEE access  \textbf{9},  128069--128094 (2021)

\bibitem{khan2019impact}
Khan, O.G.M., El-Saadany, E., Youssef, A., Shaaban, M.: Impact of electric vehicles botnets on the power grid. In: 2019 IEEE Electrical Power and Energy Conference (EPEC). pp.~1--5. IEEE (2019)

\bibitem{Lee2019acnsim}
Lee, Z.J., Johansson, D., Low, S.H.: {ACN-Sim: An open-source simulator for data-driven electric vehicle charging research}. 2019 IEEE International Conference on Communications, Control, and Computing Technologies for Smart Grids, SmartGridComm 2019 pp. 411--412 (2019). \doi{10.1109/SmartGridComm.2019.8909765}

\bibitem{Lee2019acndata}
Lee, Z.J., Li, T., Low, S.H.: {ACN-Data: Analysis and Applications of an Open EV Charging Dataset}. In: Proceedings of the Tenth ACM International Conference on Future Energy Systems. p. 139–149. e-Energy '19, Association for Computing Machinery, New York, NY, USA (2019)

\bibitem{li2016portunes+}
Li, H., D{\'a}n, G., Nahrstedt, K.: Portunes+: Privacy-preserving fast authentication for dynamic electric vehicle charging. IEEE Transactions on Smart Grid  \textbf{8}(5),  2305--2313 (2016)

\bibitem{NIPS2017_7062}
Lundberg, S.M., Lee, S.I.: A unified approach to interpreting model predictions. In: Guyon, I., Luxburg, U.V., Bengio, S., Wallach, H., Fergus, R., Vishwanathan, S., Garnett, R. (eds.) Advances in Neural Information Processing Systems 30, pp. 4765--4774. Curran Associates, Inc. (2017), \url{http://papers.nips.cc/paper/7062-a-unified-approach-to-interpreting-model-predictions.pdf}

\bibitem{marchiori2023your}
Marchiori, F., Conti, M.: Your battery is a blast! safeguarding against counterfeit batteries with authentication. In: Proceedings of the 2023 ACM SIGSAC Conference on Computer and Communications Security. pp. 105--119 (2023)

\bibitem{marchiori2025leaky}
Marchiori, F., Conti, M.: Leaky batteries: A novel set of side-channel attacks on electric vehicles. arXiv preprint arXiv:2503.08956  (2025)

\bibitem{mookherji2024secure}
Mookherji, S., Odelu, V., Prasath, R.: Secure ultra fast authentication protocol for electric vehicle charging. Computers and Electrical Engineering  \textbf{119},  109512 (2024)

\bibitem{muhammad2023emerging}
Muhammad, Z., Anwar, Z., Saleem, B., Shahid, J.: Emerging cybersecurity and privacy threats to electric vehicles and their impact on human and environmental sustainability. Energies  \textbf{16}(3), ~1113 (2023)

\bibitem{multin2018iso}
Mültin, M.: Iso 15118 as the enabler of vehicle-to-grid applications. In: 2018 International Conference of Electrical and Electronic Technologies for Automotive. pp.~1--6 (2018)

\bibitem{nasr2023chargeprint}
Nasr, T., Torabi, S., Bou-Harb, E., Fachkha, C., Assi, C.: Chargeprint: A framework for internet-scale discovery and security analysis of ev charging management systems. In: NDSS (2023)

\bibitem{shen2012charging}
Shen, W., Vo, T.T., Kapoor, A.: Charging algorithms of lithium-ion batteries: An overview. In: 2012 7th IEEE conference on industrial electronics and applications (ICIEA). pp. 1567--1572. IEEE (2012)

\bibitem{thundersky}
ThunderSky: Instruction manual for {LFP}/{LCP}/{LMP} lithium power battery. Tech. rep., Thunder Sky (2007)

\bibitem{atmSkimming}
{United States Secret Service}: {ATM \& POS Skimming}. \url{https://www.secretservice.gov/investigations/skimming} (2024)

\bibitem{unterweger2022analysis}
Unterweger, A., Knirsch, F., Engel, D., Musikhina, D., Alyousef, A., de~Meer, H.: An analysis of privacy preservation in electric vehicle charging. Energy Informatics  \textbf{5}(1), ~3 (2022)

\bibitem{ye2020cyber}
Ye, J., Guo, L., Yang, B., Li, F., Du, L., Guan, L., Song, W.: Cyber--physical security of powertrain systems in modern electric vehicles: Vulnerabilities, challenges, and future visions. IEEE Journal of Emerging and Selected Topics in Power Electronics  \textbf{9}(4),  4639--4657 (2020)

\bibitem{zhang2016privacy}
Zhang, H., Shu, Y., Cheng, P., Chen, J.: Privacy and performance trade-off in cyber-physical systems. IEEE Network  \textbf{30}(2),  62--66 (2016)

\bibitem{zhang2023realistic}
Zhang, J., Wang, Y., Jiang, B., He, H., Huang, S., Wang, C., Zhang, Y., Han, X., Guo, D., He, G., et~al.: Realistic fault detection of li-ion battery via dynamical deep learning. Nature Communications  \textbf{14}(1), ~5940 (2023)

\end{thebibliography}

\appendix
\section{Appendix}

\subsection{Hyperparameters}\label{app:hyperparams}

Table~\ref{tab:models} lists the models employed in classification and their hyperparameters. 

\begin{table}[!htpb]
  \centering
  \footnotesize
  \caption{Hyperparameters employed in Grid Search.}
  \label{tab:models}
  \renewcommand{\arraystretch}{1.1} %
  \begin{tabular}{l|l}
    \hline
    \textbf{Models} & \textbf{Hyperparameters} \\
    \hline
    \multirow{1}{*}{AdaBoost (AB)} & $\bullet$ Number of estimators \\
    \hline
    \multirow{2}{*}{Decision Tree (DT)} & $\bullet$ Criterion \\
                                        & $\bullet$ Maximum Depth \\
    \hline
    \multirow{2}{*}{k-Nearest Neighbors (kNN)} & $\bullet$ Number of neighbors \\
                                                                  & $\bullet$ Weight function \\
    \hline
    \multirow{3}{*}{Neural Network (NN)} & $\bullet$ Hidden layer sizes \\
                                         & $\bullet$ Activation function \\
                                         & $\bullet$ Solver \\
    \hline
    \multirow{2}{*}{Random Forest (RF)} & $\bullet$ Criterion \\
                                                           & $\bullet$ Number of estimators \\
    \hline
  \end{tabular}
\end{table}

\subsection{Feature Importance}\label{app:features}

The 10 most important features for the classification of the legitimate \ac{ev} are the following:
\begin{enumerate}  
    \item \textit{FFT angle (coeff. 15)} -- The phase angle of the 15th Fourier coefficient, capturing periodic patterns in the voltage signal.  
    \item \textit{FFT angle (coeff. 99)} -- Similar to the previous feature but for the 99th Fourier coefficient, highlighting high-frequency behaviors.  
    \item \textit{FFT angle (coeff. 7)} -- The phase angle of the 7th Fourier coefficient, indicative of lower frequency components.  
    \item \textit{Max Langevin point} -- A stability indicator derived from stochastic differential equations, estimating system behavior.  
    \item \textit{Change quantiles (mean)} -- Measures the mean absolute change between specific quantiles, capturing shifts in voltage distribution.  
    \item \textit{Friedrich coefficient (2)} -- A coefficient from Friedrich's method, modeling the dynamics of the voltage signal.  
    \item \textit{Mean absolute change} -- Computes the mean of absolute differences between consecutive voltage values, indicating signal variability. 
    \item \textit{Partial autocorrelation (lag 2)} -- Measures the correlation of the voltage signal with its past values at a lag of 2, detecting dependencies. 
    \item \textit{Standard deviation} -- Quantifies the overall dispersion and variation of the voltage signal.  
    \item \textit{Wavelet coeff. (w=2, coeff. 4)} -- Extracted from continuous wavelet transform, capturing multi-scale fluctuations in the voltage signal.  
\end{enumerate}

\end{document}